\documentclass[aps,prl,floatfix,superscriptaddress,twocolumn,footinbib]{revtex4-1}

\usepackage{amssymb}
\usepackage{amsmath}
\usepackage{amsfonts}
\usepackage{appendix}
\usepackage{bm}
\usepackage{graphicx}
\usepackage{epsfig}
\usepackage{epstopdf}
\usepackage{balance}
\usepackage[dvipsnames]{xcolor}
\usepackage{calc}
\usepackage{natbib}
\usepackage[colorlinks,
            linkcolor=blue,
            anchorcolor=blue,
            citecolor=blue,
            urlcolor=blue]{hyperref}
\usepackage{lipsum}

\begin{document}
\title{Flat band localization due to self-trapped orbital}
\author{Zhen Ma}\thanks{These authors contributed equally to this work.}%
\affiliation{School of Physics and Wuhan National High Magnetic Field Center,
Huazhong University of Science and Technology, Wuhan 430074,  China}
\author{Wei-Jin Chen}\thanks{These authors contributed equally to this work.}
\affiliation{School of Optical and Electronic Information, Huazhong University
of Science and Technology, Wuhan 430074,  China}
\author{Yuntian Chen}
\email{yuntian@hust.edu.cn}
\affiliation{School of Optical and Electronic Information, Huazhong University
of Science and Technology, Wuhan 430074,  China}
\author{Jin-Hua Gao}
\email{jinhua@hust.edu.cn}
\affiliation{School of Physics and Wuhan National High Magnetic Field Center,
Huazhong University of Science and Technology, Wuhan 430074,  China}
\author{X. C. Xie}
\affiliation{International Center for Quantum Materials, School of Physics, Peking University, Beijing 100871, China }
\affiliation{Collaborative Innovation Center of Quantum Matter, Beijing 100871, China}
\affiliation{CAS Center for Excellence in Topological Quantum Computation, University of Chinese Academy of Sciences, Beijing 100190, China}
\begin{abstract}
We discover a new wave localization mechanism in a periodic system without any disorder, which can  produce a novel type of perfect flat band and  is distinct from the known localization mechanisms, \textit{i}.\textit{e}., Anderson localization and flat band lattices. The first example we give is  a designed electron waveguide on  2DEG with special periodic confinement potential. Numerical calculations show that, with proper confinement geometry, electrons can be completely localized in an open waveguide. We interpret this flat band localization phenomenon by introducing the concept of self-trapped orbitals. In our treatment, each unit cell of the waveguide is equivalent to an artificial atom, where the self-trapped orbital is one of its eigenstates with unique spatial distribution.  These self-trapped orbitals form the flat bands in the waveguide.  This flat band localization through self-trapped orbitals  is a general phenomenon of wave motion, which can arise in any wave systems with carefully engineered boundary conditions. We then design a metallic waveguide array to illustrate that similar flat band localization can be readily realized and observed with electromagnetic waves.
\end{abstract}

\maketitle
Localization is an interesting phenomenon of waves. The celebrated example is the Anderson localization, where free electrons (or classical wave) can be localized by the disorder induced interference\cite{anderson1958,today2009}. In some  flat band lattices,  free electrons can be localized without any disorder, and form a flat band\cite{lattice1986,lieb89,Mielke1991, *Mielke1991b,tasaki92,2018review}. This is the  flat band localization, which results from the lattice geometry induced destructive interference of the electron wave.

The flat band localization is  of special interest and has  been intensively studied in last three decades.  It was first noticed in electron lattice system\cite{lattice1986}. Due to its quenched kinetic energy, it is predicted that the flat band electrons can be changed into ordered states (\textit{e}.\textit{g}. SDW )  by tiny Coulomb interaction\cite{lieb89,Mielke1991,Mielke1991b,tasaki92,wu2007,sc2007}.  Meanwhile,  if the flat band is topologically nontrivial,  the fractional quantum hall states can be realized in the flat band lattices in the absence of magnetic field\cite{wen2011,sun2011,mudry2011}.
Recently, it is proposed that the flat band localization can be realized in some artificial lattice systems\cite{2018review},  $\textit{e}.\textit{g}.$  artificial electron lattice on  metal surface\cite{lieb2016,*ma2017,slot2017,atomlattice2017}, cold atoms in optical lattice\cite{xing2010,pra2010,Taiee1500854,liebprl2017}, photonic lattices\cite{njp2014,thomson2015,rodrigo2015},  cavity QED systems\cite{huyong2016}, optomechanical arrays\cite{Wanoe2017,*Wan2018}, and fractal (quasi-periodic) lattices\cite{fractal2018,xie1988}. 

\begin{figure*}
\centering
\includegraphics[width=18cm]{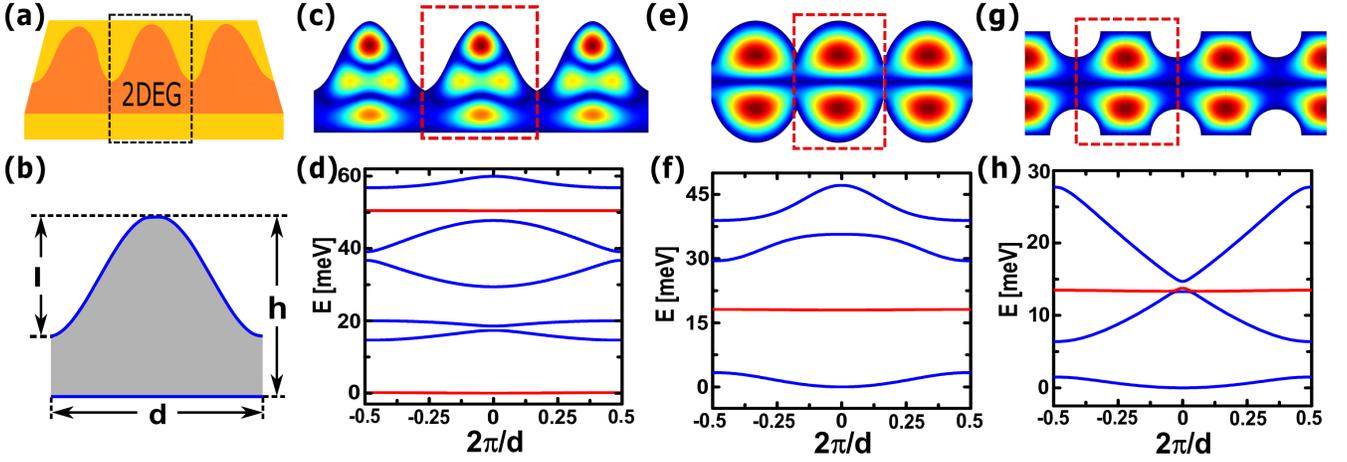}
\caption{(a) Schematic of the electron waveguide on 2DEG. The electrons are confined in a dark-colored region, and the dashed  box indicates an unit cell of the waveguide. (b) One example of $U(r)$ in one unit cell. $U(r)$ is zero inside the waveguide (the gray region) and infinite outside.  Here, $d$=52 nm, $h$=40 nm, $l$=26 nm.  The corresponding band structure is given in (d) and $|\phi_k(r)|^2$ of the flat band (upper one) at the $\Gamma$ point is plotted in (c).  (e) and (g) are  two other electron waveguides, where $U(r)$ are of different shapes, and the corresponding band structures are given in (f) and (h), respectively. $|\phi_k(r)|^2$ of the flat band ($\Gamma$ point) are also plotted in (e) and (f). The geometry parameters for (e) and (f) are given in the Supplementary Materials. }
\label{fig1}
\end{figure*}

In this work, we discover a new type of wave localization phenomenon in  periodic systems, which can produce prefect flat bands with designed band structures.  It clearly indicates that interference induced localization in periodic systems is not just restricted to the tight-binding lattices, but should be a more general phenomenon.  
 We first design  a quasi one dimensional waveguide with a periodic confinement potential, \textit{e}.\textit{g}. electron waveguides on two dimensional electron gas  (2DEG) as illustrated in Fig. \ref{fig1}.  Numerical calculations show that, if choosing proper shape of the confinement  potential, electrons can be completely localized in an open waveguide and perfect flat bands arise. We then illustrate that this unusual localization actually results from the emergence  of \textit{self-trapped orbitals}.  Specifically, due to the periodicity of confinement potential, the electron waveguide can be divided into unit cells, where each one is equivalent to an artificial atom (or a quantum dot). Correspondingly, the eigenfunctions of such atom can be viewed as the orbitals of the artificial atom. Because of the  designed shape of potential,  some peculiar orbitals have special spatial distributions (or interference patterns), which render themselves localized in  the open electron waveguide. These self-trapped orbitals give rise to flat bands in the electron waveguide. In principle, the self-trapped orbital should be viewed as a new kind of compact localized state, which was first proposed in the lattice systems\cite{cls2017,cls2018}. 
 Importantly, since that the electron filling of the 2DEG is controllable, we can tune the Fermi level of the electron waveguide to the position of flat band. Therefore, it offers an ideal platform to investigate the flat band induced correlated ground states, \textit{e}.\textit{g}.  ferromagnetism, Wigner crystal, etc. 
 
 We further point out that this flat band localization is a general phenomenon of waves, and can arise in any wave system with proper boundary condition. Taking the electromagnetic wave for example, we design a metallic waveguide array, and show that similar flat band localization can be readily observed in experiment.
The highly tuneable self-trapped electromagnetic modes are related to many intriguing optical applications, such as information  transmission\cite{ol2014,ol2016}, slow light\cite{slowlight2008} and  bound states in continuum \cite{BIC2016}.

The models we studied are illustrated in Fig. \ref{fig1}. Here, we take the 2DEG system as the first example. The  Hamiltonian is
\begin{equation}\label{H1}
H=-\frac{\hbar^2}{2m}\nabla^2+U(r).
\end{equation} 
The free electrons are confined by a  periodic infinite potential well $U(r)$ to  form an electron  waveguide, where several examples of $U(r)$ with different shape are given in Fig. \ref{fig1}.   $U(r)$ is periodic along the direction of the waveguide, so that we can define the unit cell of the electron waveguide, as indicated by the dashed boxes.    
An example of $U(r)$ (in one unit cell) is given in Fig. \ref{fig1}  (b), which corresponds to the electron waveguide in Fig. \ref{fig1} (c). $U(r)$ is zero inside the electron waveguide (grey region) and infinitely large outside. In experiments, this confinement potential on 2DEG can be achieved by applying a proper gate voltage or lithography. Note that the required parameters here should be within the  present experiment capability.
Numerically,  we use the finite difference method  to solve the Schr\"{o}dinger equation with proper boundary conditions [see Supplementary Materials (SM)]. Since it is a periodic system, we can calculate the energy bands and the Bloch wave functions.  The effective mass is $0.067$ $m_e$, which is the value of AsGa 2DEG.

We plot the energy bands for the electron waveguides with different shapes in  Fig. \ref{fig1} . In all those designed electron waveguides, we get  perfect flat bands with different band structures. It is the central result of this work. For example, with the $U(r)$ in Fig. \ref{fig1} (c), the lowest seven bands are plotted in Fig. \ref{fig1} (d) (the whole band structure is given in SM). We see that in such electron waveguide, there are two completely flat bands (red solid lines), while others are dispersive (blue solid lines). Interestingly, the flat bands are separated away from the dispersive bands with  gaps  from several to tens meV. The gaps can be further tuned by changing the size of the electron waveguide.  It means that, in such electron waveguide, we can access the flat band states without any disturbances from other states.  This offers great convenience to study the property of flat band.  To  set the Fermi level in the upper flat band, the required charge density is about $1.2\times 10^{-11}$ $cm^{-2}$, which is a reasonable value in experiment [see SM]. Actually, various kinds of flat bands can be obtained  by this method with different $U(r)$. In Fig.  \ref{fig1} (f),  single flat band emerges, where the potential $U(r)$ is given in Fig. \ref{fig1} (e). We can also get  a flat band contacted with other bands at certain k points [see in Fig. \ref{fig1} (h)], with the $U(r)$ given in Fig. \ref{fig1} (g).
The spatial distribution of the flat band wave functions $|\phi_{k}(r)|^2$ at $\Gamma$ point
are plotted in Fig. \ref{fig1} (c), (e), (g), respectively.
Obviously, this new flat band localization phenomenon in waveguide is beyond the scope of the well-known flat band lattice models , such as Lieb lattice, Kagome lattice, \textit{et} \textit{al}.  Actually, it implies that there is a geometry induced wave localization mechanism in a general periodic system. 

\begin{figure}
\centering
\includegraphics[width=8.5cm]{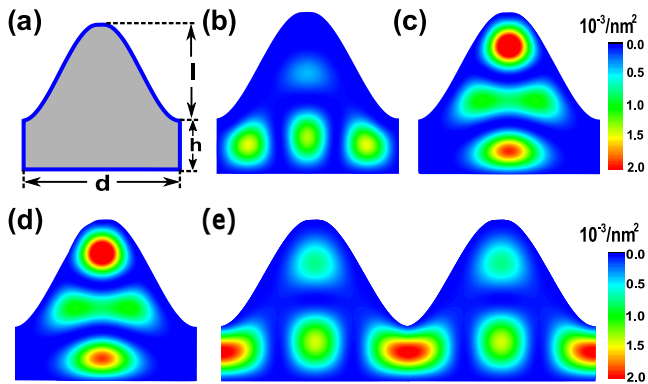}
\caption{(a) $U(r)$ of  an isolated  artificial atom (or quantum dot). The geometry parameters are the same as Fig. \ref{fig1} (b).  $U(r)$ is zero inside the atom (gray region) and infinite outside. (b) and (c) are the wave functions $|\phi(r)|^2$ of the fourth and sixth orbitals (eigenstates) in the isolated atom, respectively. (d) is the  wave function $|\phi_k(r)|^2$ of the fourth band at $\Gamma$ point for the electron waveguide in Fig. \ref{fig1} (d),  while  (e) is that of the sixth band (upper flat band).   We plot one unit cell in (d) but two unit cells in (e) in order to show the bond between two adjacent orbitals.  }
\label{fig2}
\end{figure}

To understand this exotic flat band localization phenomenon, we first introduce the concept of artificial orbital.  We take the case in Fig. \ref{fig1} (c),(d) as an typical example.  Basically, each unit cell of the electron waveguide can be  viewed as a single artificial atom (or a quantum dot), in which electrons are confined in a potential like in Fig. \ref{fig2} (a). Then, these artificial atoms (unit cells) are connected to construct an atomic chain, \textit{i}.\textit{e}. electron waveguide. Note that the potential for an isolated atom [ Fig. \ref{fig2} (a)] is different from that for the electron waveguide [Fig. \ref{fig1} (b)] in boundary condition.   For a single isolated atom, the potential is spatially closed, such that electrons in the atom have no way to escape. However, as the atoms are connected, the potential well is opened at the connecting edges [Fig. \ref{fig1} (b)], through which electrons can transport to adjacent atoms. 
In a single artificial atom, the energy of eigenstates are discrete due to the spatial confinement, and the corresponding  eigenfunctions present different interference patterns. These eigenfunctions are just the orbitals of the artificial atom.

 The artificial orbitals here are very different from the normal atomic orbitals. Importantly, the shape of the orbital (\textit{i}.\textit{e}.\ interference pattern) here is only the effect of interference, and depends on the geometry of the confinement potential. 
It is because that, unlike the Coulomb interaction, the confinement potential of the artificial atom [Fig. \ref{fig2} (a)] is  flat inside the artificial atom and has a controllable geometry. Due to the overlap between the orbitals, the energy levels of the artificial atom are transformed into the energy bands of the electron waveguide. This orbital picture works quite well for the bands of the electron waveguides  we studied here.   And we would like to mention that similar artificial electron orbitals induced by  2D confinement potential have been studied theoretically\cite{ma2017,lieb2016, Qiu2019}, and observed in  recent experiments\cite{slot2017,prx2019}.


Because of the peculiar interference patterns, some orbitals in this artificial atom are self localized (or self trapped), which result in the flat bands of the electron waveguide. In Fig. \ref{fig2} (b), we plot the wave function of the fourth artificial orbital in a single atom, which corresponds to the fourth band (dispersive) in Fig. \ref{fig1} (d). And, for the upper flat band, \textit{i}.\textit{e}., the sixth bands in Fig. \ref{fig1} (d), the corresponding artificial orbital (the sixth  orbital) is given in Fig. \ref{fig2} (c). Comparing the two  orbitals, it can be found that the flat band orbital is localized in a region far away from the connecting edges, because of its unique interference pattern (all the other artificial orbitals are given in SM).  It implies that, when atoms are connected, the  overlap between the flat band orbitals of adjacent atoms is zero, and thus electrons in this orbital actually are strictly localized in each unit cell. We name this kind of artificial orbital \textit{self-trapped orbital}.  The self-trapped orbital is the reason why the electron waveguide has flat bands. To make clear this point, we plot the Bloch wave function  at  the  $\Gamma$ point for the upper flat band in Fig. \ref{fig2} (d). Note that the self-trapped orbital of the artificial atom [Fig. \ref{fig2} (c)] and the corresponding flat band Bloch wave function in electron waveguide [Fig. \ref{fig2} (d)]  are exactly the same.   Meanwhile, for other dispersive energy bands, we can clearly see that bonds are formed in between the adjacent atoms [Fig. \ref{fig2} (e)].

\begin{figure*}
\centering
\includegraphics[width=18cm]{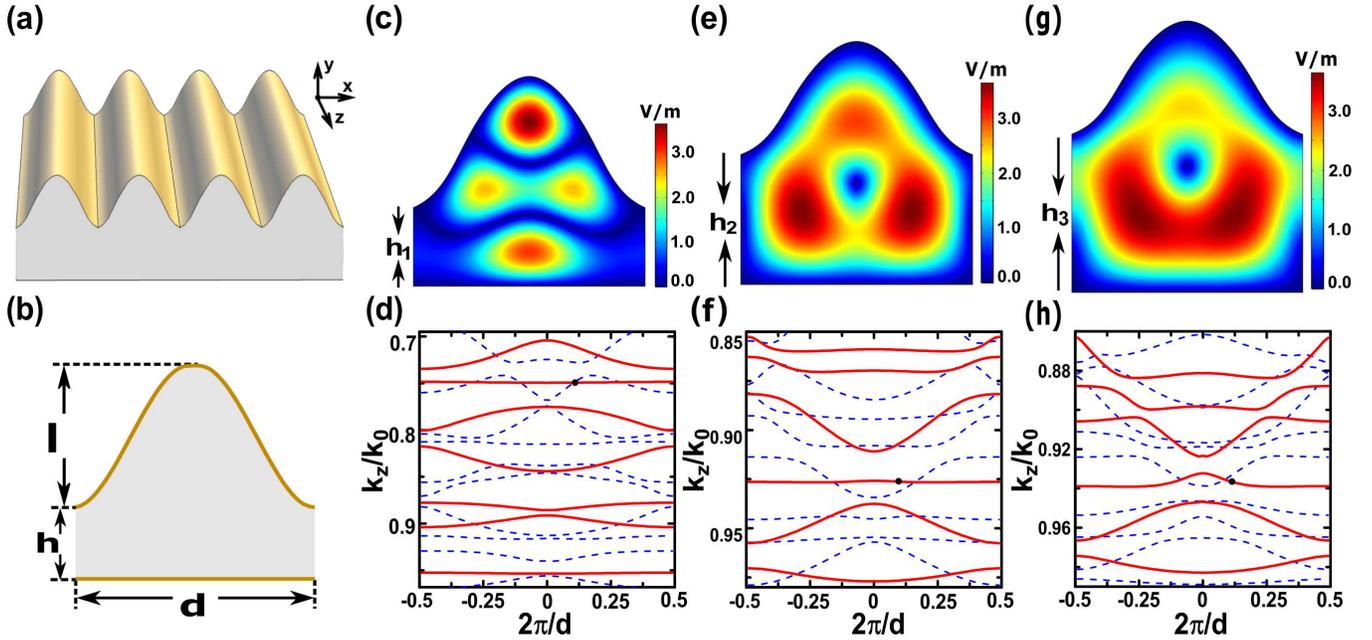}
\caption{(a) Schematic of the metallic waveguide array, where the shape of the cross section of one unit cell in x-y plane is given in (b). In (b), the geometry of each unit cell is determined by d,l,h, where d is the lattice constant, h is the height between the bottom and the lowest upper  surface, l is the hight of the bump of the upper surface. The electromagnetic wave propagate along z-axis, and can leak from one waveguide to the neighbouring waveguide  in the x-y plane; Three different waveguide lattices, i.e., labelled as w1, w2, w3, are examined, see the mode profiles ($E_z$)  in (c/e/g) and the band diagram in (g/f/g) for w1,w2,w3. The geomentric parameters for  w1, w2, w3 waveguide lattice are given by as follows,  $d/l/h$ = 16/8/5 cm for w1, $d/l/h$ =16/8/9.2 cm for w2, $d/l/h$ = 16/8/11.2 cm for w3.}
\label{fig3}
\end{figure*}

The self-trapped orbital is a pure interference phenomenon. 
We would like to emphasize that, the self-trapped orbital induced flat bands here are different from  the trivial flat bands, which result from the potential barrier. An example of trivial  flat band is the core electrons in real atoms. In real atoms, the energy of core electrons are lower than that of valence electrons, such that the core electrons are tightly bound to the nucleus due to the confinement of Coulomb potential. The wave function of the core electron are very localized, and the potential barrier does not allow the core electrons to move from one atom to another. This produces a trivial flat band for the core electrons, which results from the potential barrier rather than the wave interference. In our model, the potential inside the electron waveguide is flat, such that there is no potential barrier to obstruct the moving of electrons inside the electron waveguide. Thus, the self localization induced flat band is an interference phenomenon, and is different from the trivial flat band case. In addition, in some sense, the self-trapped orbital here  is very similar to the compact localized state in flat band lattice\cite{cls2017}, which is the localized eigenstate of a periodic system.

The  self-trapped orbitals, as well as the corresponding flat bands,  are controlled by the local geometry of the $U(r)$. With different local geometry of $U(r)$, we can get various self-trapped orbitals, and thus different flat band structures. This is shown clearly in Fig. \ref{fig1}, where various flat band structures have been obtained via choosing different shape of $U(r)$.  Note that, the energy  of the flat bands and their relative positions to other dispersive bands are all controllable by designing a proper $U(r)$. This  property is very useful for the future application of the flat bands. Meanwhile, the flatness of the flat band can also be tuned by the shape of $U(r)$. Fig. \ref{fig1} shows that strictly flat energy bands, i.e. completely localized electrons, can be realized in the designed electron waveguides. Actually, there are still some fundamental difficulties in obtaining perfect electron flat bands in experiment in the TB lattice models. This is because that, in real systems, long range hopping always exist. Thus, the flat band lattices normally will provide some bent flat bands in experiment\cite{lieb2016} . In contrast,  there is no such difficulty  in our scheme. Meanwhile, the self-trapped orbitals and the flat bands are stable against the small potential disturbance as along as it does not drastically modify the spaticial distribution of the wave function. 

So far, all the discussions are about the electron system. Again, we emphasize that the  flat band localization proposed here can generally arise in any wave system due to its generic nature of wave interference.  To illustrate this idea, we give the second example of self-trapped orbital induced flat band in  electromagnetic wave system. Concretely, we show that the similar flat band localization  of electromagnetic waves can be realized in  waveguide arrays, see  Fig. \ref{fig3} (a). The metallic waveguide array is infinitely long along the z direction, and the cross-section  in x-y plane has a similar shape as that in Fig. \ref{fig1} (c). Here, we consider the TE modes where the electric field has only one component along z-axis, as given by $E_z(x,y,z)=E_z(x,y)\exp(-i k_z z)$. Accordingly, $E_z(x,y)$ in the waveguide obeys the equation as follows,
\begin{equation}\label{waveguide}
[\nabla^2 +k^2_0] E_z=k^2_0 n^2_{eff}E_z
\end{equation}
where $k_0=2\pi / \lambda_0$, $n_{eff}=k_z/k_0$, and $\lambda_0$  is vacuum wavelength. Here, $n_{eff}$ can be determined by the eigenvalue of this equation, and $n_{eff}^2$ corresponds to the energy in Schr\"{o}dinger equation. For the metallic waveguide,  the boundary condition is $E_z=0$.  The Eq. \eqref{waveguide} is  analogous to the Schr\"{o}dinger equation, and the shape of the metallic waveguide mimics the confinement potential $U(r)$ in the electron waveguide in Fig.\ \ref{fig1} (c). Thus, it can be expected that TE modes here have similar behaviours as the electrons in  the electron waveguide. Note that there are also TM modes in the waveguide, which is not discussed. The band structure of the metallic waveguide array is calculated using a commercial software package: COMSOL MULTIPHYSICS.

As expected, the numerical results unambiguously show that the flat bands of electromagnetic wave also exist in the coupled waveguide array, as given in Fig. \ref{fig3}. In order to facilitate the comparison, we plot three cases, where  the waveguide geometry parameters $l$, $d$ are fixed, and different values of $h$ are  used, as sketched in Fig. \ref{fig3} (b). In Fig. \ref{fig3} (c), (e), (g), we set $h=5, 9.2, 11.2$ $cm$, respectively. And the  band diagrams are given in Fig. \ref{fig3} (d), (f), (h) accordingly. All the bands for TE (TM) modes are plotted as red solid (blue dashed) lines in Fig. \ref{fig3} (d-h) . 
In Fig. \ref{fig3} (c) and (d),  the TE modes (see the red solid lines)  have very similar  band structure as that of the electron waveguide in Fig. \ref{fig1} (d). The eigen-field at the $\Gamma$ point of the flat band is  plotted in Fig. \ref{fig3} (c),  which also has a similar mode profile as the flat band wave function in electron waveguide. The flat bands indicate that now the corresponding TE mode is localized along the x direction due to the self-localized phenomenon. When we change the geometry of confinement, the artificial orbitals are changed and we get different band structure.  In Fig. \ref{fig3} (e) and (f), we enlarge $h$ to $9.2$ $cm$ and the band structure is fundamentally changed. Interestingly, the third band becomes flat here, while the former flat bands in Fig. \ref{fig3} (d) are changed to be dispersive. We also plot the eigen-field of the flat band in Fig. \ref{fig3} (e), and we see a rather different self-trapped orbital. If the height $h$ becomes larger than a critical value, e.g. $h=11.2$ cm in Fig. \ref{fig3} (g) and (h),  all the bands are dispersive and the localization phenomenon vanishes. Finally, we emphasize that,  though the designed waveguide array above is for the microwave region, similar phenomenon can occur for the electromagnetic waves with scaled wavelength.

In summary, we have shown that the flat band localization of wave can occur not only in the TB lattices as was well known, but also in a more general periodic system without any disorder.  Using  electron waveguide on 2DEG and metallic waveguide array as  examples, we numerically demonstrate that this kind of flat band localization is a general property of wave (both quantum and classical waves).  We give an intuitive understanding about this new wave localization phenomenon by introducing the concept of self-trapped orbitals. Though the examples are quasi one dimension, we argue that similar phenomenon can also arise in two and three dimension. 
At last, our study about the self-trapped orbitals (modes)  here once again illustrates that the orbital is an important concept in artificial periodic systems\cite{ma2017,Qiu2019}, and can give rise to  exotic phenomena that do not exist in real materials.

\begin{acknowledgments}
We thank the supports by the National Natural Science Foundation of China (Grants No. 11534001, 11874160, 11274129, 11874026， 61405067), and the Fundamental Research Funds for the Central Universities (HUST: 2017KFYXJJ027), and NBRPC (Grants No. 2015CB921102).
\end{acknowledgments}


\bibliography{flatbandlocalization}

\end{document}